\documentclass[12pt]{iopart}

\pdfoutput=1

\usepackage{iopams}
\usepackage{dcolumn}
\usepackage{graphicx}
\usepackage{bm}

\newcommand{\AFA}{$A$Fe$_{2}$As$_{2}$}
\newcommand{\BFA}{BaFe$_{2}$As$_{2}$}
\newcommand{\SFA}{SrFe$_{2}$As$_{2}$}
\newcommand{\EFA}{EuFe$_{2}$As$_{2}$}

\newcommand{\BKSC}{(Ba$_{0.6}$K$_{0.4}$)Fe$_{2}$As$_{2}$}
\newcommand{\TCS}{ThCr$_{2}$Si$_{2}$}

\begin{document}

\rapid[Phase transitions in \SFA\ and \EFA]{Structural and magnetic phase transitions in the ternary iron arsenides \SFA\ and \EFA}

%


\author{Marcus Tegel$^1$, Marianne Rotter$^1$, Veronika Weiss$^1$, Falko M. Schappacher$^2$, Rainer P\"ottgen$^2$ and Dirk Johrendt$^1$}

\address{$^1$Department Chemie und Biochemie der Ludwig-Maximilians-Universit\"{a}t M\"{u}nchen, Butenandtstr- 5-13 (Haus D), 81377 M\"{u}nchen, Germany}
\address{$^2$Institut f\"{u}r Anorganische und Analytische Chemie, Universit\"{a}t M\"{u}nster, Corrensstrasse 30,
D-48149 M\"{u}nster, Germany}

\ead{johrendt@lmu.de}

\begin{abstract}
The structural and magnetic phase transitions of the ternary iron arsenides \SFA\ and \EFA\ were studied by temperature-dependent x-ray powder diffraction and $^{57}$Fe M\"ossbauer spectroscopy. Both compounds crystallize in the tetragonal \TCS-type structure at room temperature and exhibit displacive structural transitions at 203 K (\SFA) or 190 K (\EFA) to orthorhombic lattice symmetry in agreement with the group-subgroup relationship between $I4/mmm$ and $Fmmm$. $^{57}$Fe M\"{o}ssbauer spectroscopy experiments with \SFA\ show full hyperfine field splitting below the phase transition temperature (8.91(1) T at 4.2 K). Order parameters were extracted from detailed measurements of the lattice parameters and fitted to a simple power law. We find a relation between the critical exponents and the transition temperatures for \AFA\ compounds, which shows that the transition of \BFA\ is indeed more continuous than the transition of \SFA\, but it remains second order even in the latter case.
\end{abstract}


\pacs{
 61.50.Ks,
 74.10.+v,
 33.45.+x 
 }

\vspace{2pc}
\noindent{\it Keywords}: Iron arsenides, Phase transition, Structure, $^{57}$Fe M\"ossbauer spectroscopy\\

\maketitle

\section{Introduction}

The discovery of the iron arsenide superconductors has provided fresh impetus in the field of high-$T_C$ superconductivity \cite{Hosono-2008, Angew-2008}. LaFeAsO with the tetragonal ZrCuSiAs-type structure \cite{Jeitschko-1974} becomes superconducting by doping the (FeAs)$^{\delta-}$ layers either with electrons or holes \cite{LaFeAsOx-41K, LaSrFeAsx-25K}. The initially reported transition temperature of 26 K rose quickly up to 55 K by replacing La$^{3+}$ ions for smaller Sm$^{3+}$ ions \cite{Sm-TC55}. The crystal structure of LaFeAsO contains alternating layers of edge-sharing La$_{4/4}$O and FeAs$_{4/4}$ tetrahedra and superconductivity emerges in the FeAs layers by adding or removing about 0.2 electrons per formula unit.

\begin{figure}[h]
\center{
\includegraphics[width=0.7\textwidth]{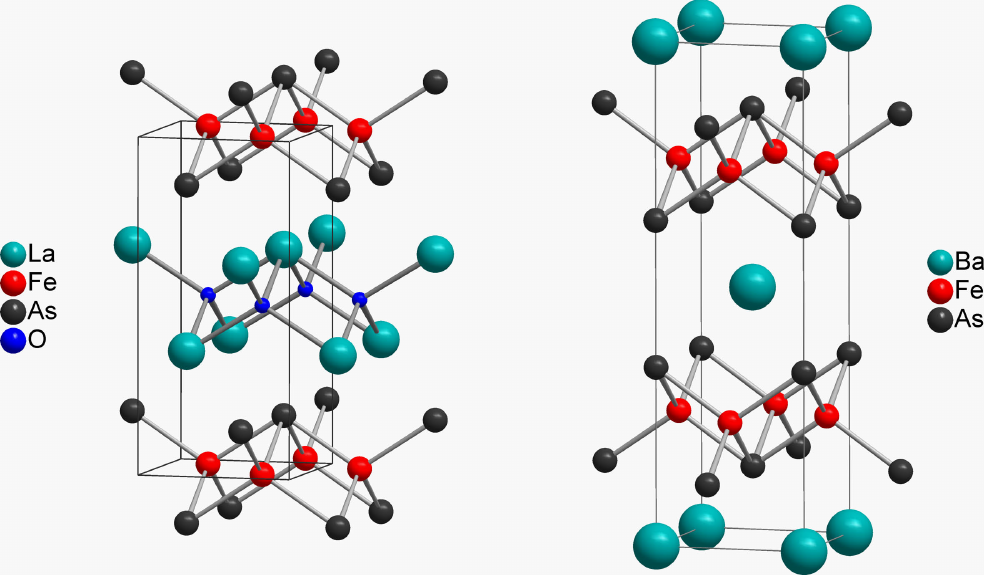}
\caption{\label{fig:Strukturen} (Color online) Crystal structures of LaFeAsO and \BFA.}
}
\end{figure}

Very recently, we reported on the oxygen-free iron arsenide \BFA\ with the \TCS-type structure as another possible parent compound for superconductivity \cite{BFA}. Soon after that we were able to induce superconductivity by hole doping in the compound \BKSC\ with $T_C$ = 38 K and we have therefore established a further family of iron arsenide superconductors \cite{BKFA}. Another report on superconductivity at 37 K in isostructural K- and Cs- doped \SFA\ followed quickly \cite{Lorenz-2008}. The crystal structures of both LaFeAsO and \BFA\ are depicted in Figure~\ref{fig:Strukturen}. Both compounds are built up by almost identical (FeAs)$^{\delta-}$ layers, but they are separated by lanthanum oxide sheets in LaFeAsO and by barium atoms in \BFA\, respectively.

The non-superconducting parent materials LaFeAsO and \BFA\ show remarkably similar properties. Both compounds are poor metals and only weakly magnetic. One key finding is the existence of a spin-density-wave (SDW) anomaly, which occurs at $T_{tr}$ = 150 K in LaFeAsO and at $T_{tr}$ = 140 K in \BFA, respectively \cite{Cruz-Neutrons, BFA}. This SDW is linked to abrupt changes in the electrical resistivity and magnetic susceptibility and also to structural phase transitions. Antiferromagnetic ordering was found in LaFeAsO 18 K  below the structural transition \cite{Cruz-Neutrons}, but directly at or at least very close to the lattice distortion temperature in \BFA\ according to recent neutron diffration experiments \cite{Huang-2008-BFA-Neutron}. It is currently believed that superconductivity in the iron arsenides is intimately connected with the suppression of this SDW anomaly by doping. This suggests that spin-fluctuations may play an important role for the mechanism of superconductivity as it was also assumed for the high-$T_C$ cuprates. Thus, the nature of the phase transitions is important for a deeper understanding of superconductivity in the iron arsenides. However, precise structural data close to the phase transition are only available for \BFA\ and LaFeAsO \cite{BFA, Nomura-2008, McGuire-2008}. \SFA\ has been studied by single crystal data with relatively low resolution \cite{Yan-2008}, which allow no evaluation of the order parameter close to the transition temperature. Furthermore, the connection of the structural transition in \SFA\ with magnetic ordering, as well as the structure of \EFA\ at low temperatures has not been investigated yet. We have therefore studied the structural phase transitions of polycrystalline \SFA\ and \EFA\ in detail by temperature-dependent x-ray powder diffraction. We could also confirm the association of the structural transition in \SFA\ with magnetic ordering by $^{57}$Fe M\"ossbauer spectroscopy.

\section{Experimental}


\SFA\ and \EFA\ were synthesized by heating mixtures of distilled Sr(Eu)-metal, iron-powder and sublimed arsenic at ratios of 1:2:2 in alumina crucibles, which were sealed in silica tubes under an atmosphere of purified argon. The mixtures were heated to 850 K at a rate of 50 K/h, kept at this temperature for 15 h and cooled down to room temperature. The reaction product was homogenized in an agate mortar and annealed at 900 K for 15 h. The obtained black crystalline powders of \SFA\ and \EFA\ are sensitive to air and moisture.


Temperature dependent x-ray powder diffraction data were collected using a \textsc{Huber} G670 Guinier imaging plate diffractometer (Cu-$K_{\alpha_{1}}$ radiation, Ge-111 monochromator), equipped with a closed-cycle He-cryostat. Rietveld refinements were performed with the \textsc{GSAS} package \cite{GSAS} using Thompson-Cox-Hastings functions \cite{Thompson-Cox-Hastings} with asymmetry corrections as reflection profiles.\cite{Finger-Cox-Jephcoat}


A $^{57}$Co/Rh source was available for the $^{57}$Fe M\"{o}ssbauer spectroscopy investigations. The velocity was calibrated relative to the signal of $\alpha$-Fe. A \SFA\ sample was placed in a thin-walled PVC container at a thickness of about 4 mg Fe/cm$^2$. The measurements were performed in the usual transmission geometry at 298, 77 and 4.2 K. The source was kept at room temperature.

\section{Results and Discussion}

\begin{figure}[h]
\center{
\includegraphics[width=0.5\textwidth]{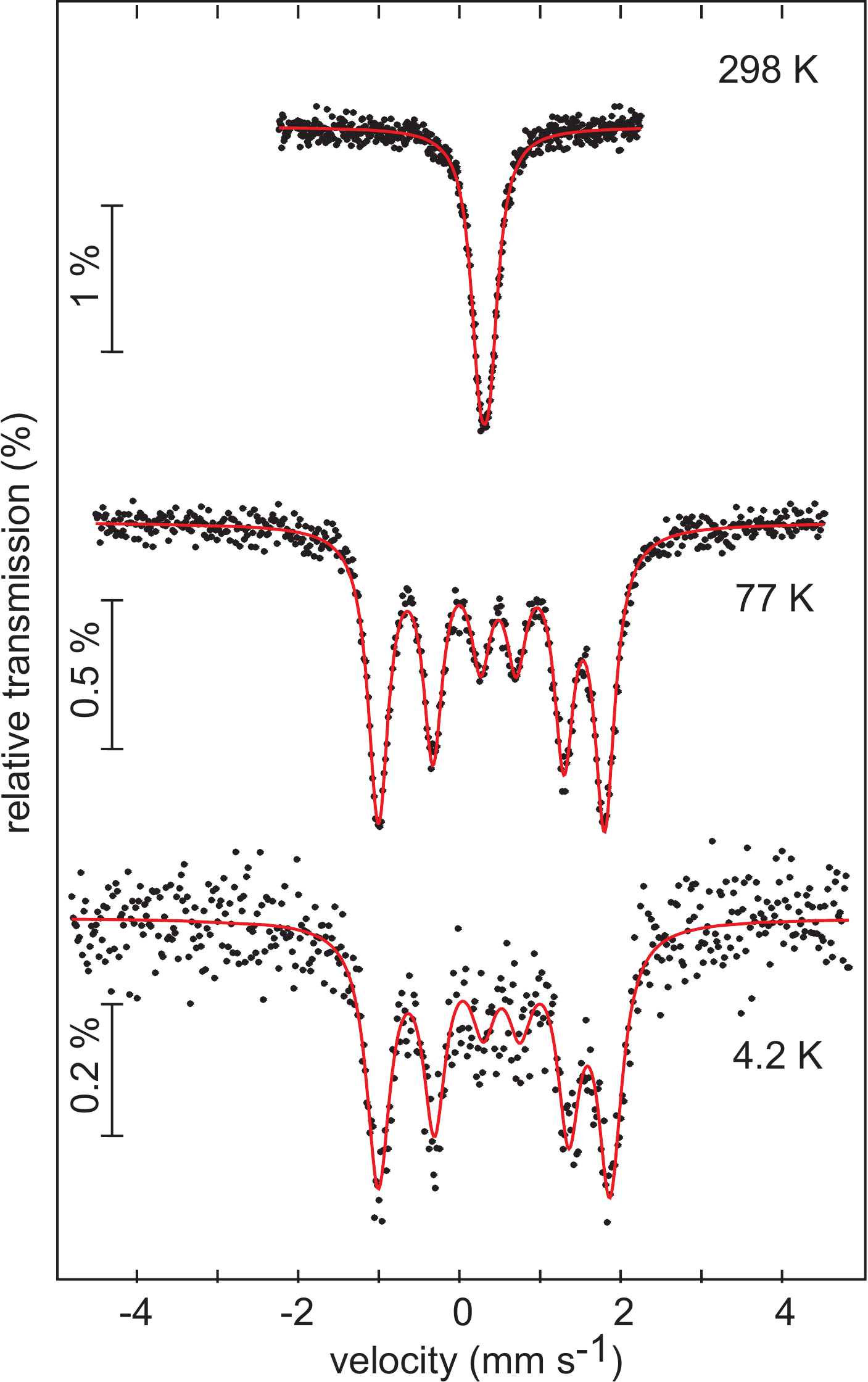}
\caption{\label{fig:Moessbauer} (Color online) $^{57}$Fe M\"ossbauer spectra of \SFA\ with transmission integral fits.}
}
\end{figure}

In order to clarify the connection of the structural phase transition in \SFA\ with magnetic ordering, we first present $^{57}$Fe M\"ossbauer spectra of \SFA\ measured at 298, 77 and 4.2 K in Figure~\ref{fig:Moessbauer} together with transmission integral fits. In agreement with the \TCS-type crystal structure we observe a single absorption line for \SFA. At 77 K, well below the structural transition temperature, we detect full magnetic hyperfine splitting of the signal. Excellent fits of the data are obtained with the parameters listed in Table~\ref{tab:MB-Data}. The isomer shifts are similar to those found in \BFA\ ($\delta$= 0.31 - 0.44 mm/s). Due to different ionic radii, we observed a smaller c/a ratio of 3.15 for \SFA\, in comparison to c/a = 3.29 for \BFA\ \cite{Nagorsen-1980}. The stronger compression of the FeAs$_{4/4}$ tetrahedra in the strontium compound is also reflected by the larger quadrupole splitting parameter. Good agreement is observed with the recently published $^{57}$Fe data for LaFePO \cite{Tegel-2008} and LaFeAsO \cite{Kitao-2008, Klauss-2008}, which contain electronically very similar tetrahedral FeP$_{4/4}$ and FeAs$_{4/4}$ layers.
The hyperfine field detected at the iron nuclei in \SFA\ ($B_{hf}$ = 8.91(1) T) at 4.2 K is considerably higher than in \BFA\ (5.47 T) \cite{BFA}. The magnetic behavior of the iron arsenide layers strongly depends on the occupation of the Fe $3d_{x^2-y^2}$ orbitals, and the latter depends on the position of the arsenic atoms \cite{Krellner-2008}. Thus, with smaller strontium and europium atoms, a stronger magnetic character of the iron arsenide layers and consequently a higher ordering temperature can be observed, i.e. 140 K in \BFA, 205 K in \SFA\ \cite{Krellner-2008}, and 200 K in \EFA\ \cite{Raffius-1993}. The hyperfine fields show the same trend: 5.47 T in \BFA\ \cite{BFA}, 8.91 T K in \SFA, and 8.5 T in \EFA\ \cite{Raffius-1993}.

\begin{table}[h]
\caption{\label{tab:MB-Data}Fitting parameters of $^{57}$Fe M\"{o}ssbauer spectroscopy measurements with \SFA. $\delta$:~isomer shift; $\Delta E_Q$:~quadrupole splitting parameter~$\Gamma$:~experimental line width; $B_{hf}$:~magnetic hyperfine field.}

\begin{center}
\begin{tabular}{lllll}
\hline
$T$ (K) &  $\delta$ (mm$\cdot$s$^{-1}$) & $\Gamma$ (mm$\cdot$s$^{-1}$) & $\Delta E_Q$ (mm$\cdot$s$^{-1}$) &  $B_{hf}$ (T)\\
\hline
298 & 0.31(1)  &  0.28(1) &  $-0.13(1)$ &  -- \\
77  & 0.44(1)  &  0.31(1) &  $-0.09(1)$ & 8.70(1)\\
4.2 & 0.47(1)  &  0.37(6) &  $-0.09(1)$ & 8.91(1)\\
\hline
\end{tabular}
\end{center}
\end{table}

\begin{figure}[h]
\center{
\includegraphics[width=0.7\textwidth]{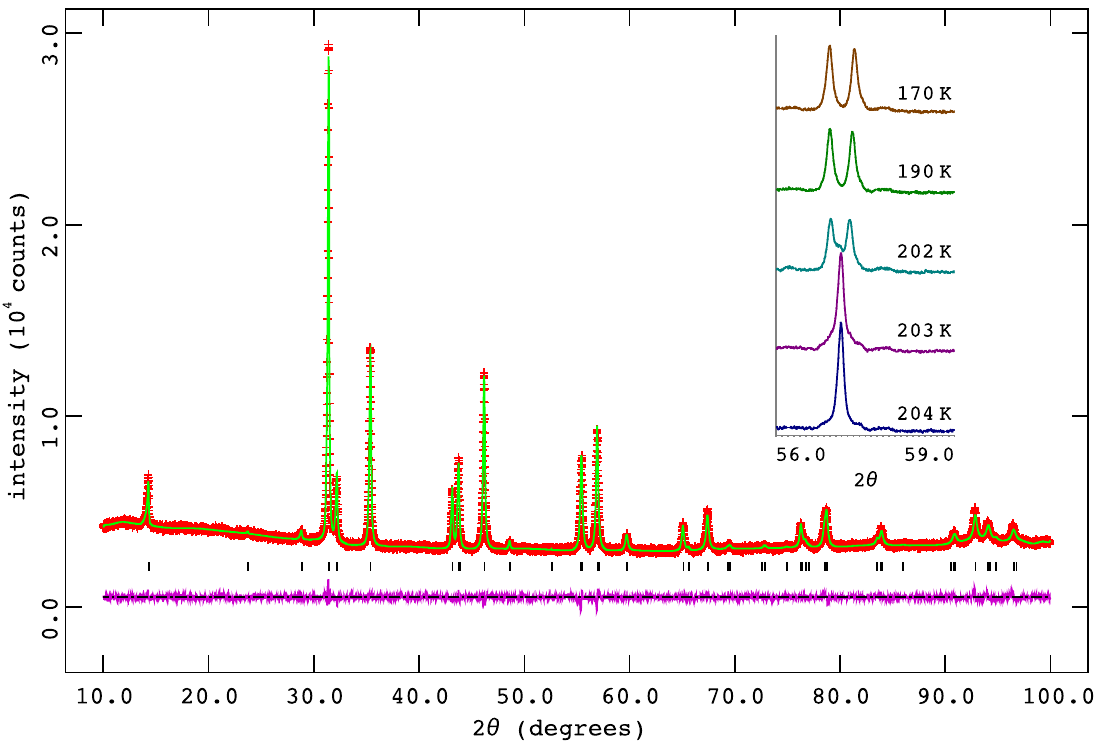}
\caption{\label{fig:SrFe2As2_X} (Color online) X-ray powder pattern measured at 297 K (+) and Rietveld fit (-) of \SFA. Insert: Splitting of the (2 1 3) reflection.}
}
\end{figure}

According to recently published single crystal data \cite{Yan-2008}, \SFA\ exhibits a structural transition from the tetragonal space group $I4/mmm$ to the orthorhombic subgroup $Fmmm$ as first described for \BFA\ \cite{BFA}. Anomalies of the physical properties have been reported for polycrystalline \SFA\ at 205 K \cite{Krellner-2008} and at 198 K for single crystals \cite{Yan-2008}. We confirm a structural transition at 203 K by x-ray powder patterns recorded at low temperatures. Figure~\ref{fig:SrFe2As2_X} shows the experimental and fitted powder pattern of \SFA\ with a clear splitting of the (2 1 3)-reflection depicted in the insert. We have refined the structures above and well below the transition temperature and obtained the crystallographic data summarized in Tables~\ref{tab:CrystallographicLT} and \ref{tab:CrystallographicRT}.

\vspace{2cm}

\vspace{2cm}

An intensively discussed question is whether the transition is of first or second order. For second order transitions, the space groups of the distorted and undistorted structures have to comply with a group-subgroup relationship according to Hermann's theorem \cite{Hermann-1929, Deonarine-1983}. The space group $Fmmm$ is a translation equivalent subgroup of $I4/mmm$ of index 2. Thus, from a group theoretical standpoint one could expect a second order transition with continuous variation of the order parameter.

\begin{figure}[h]
\center{
\includegraphics[width=0.7\textwidth]{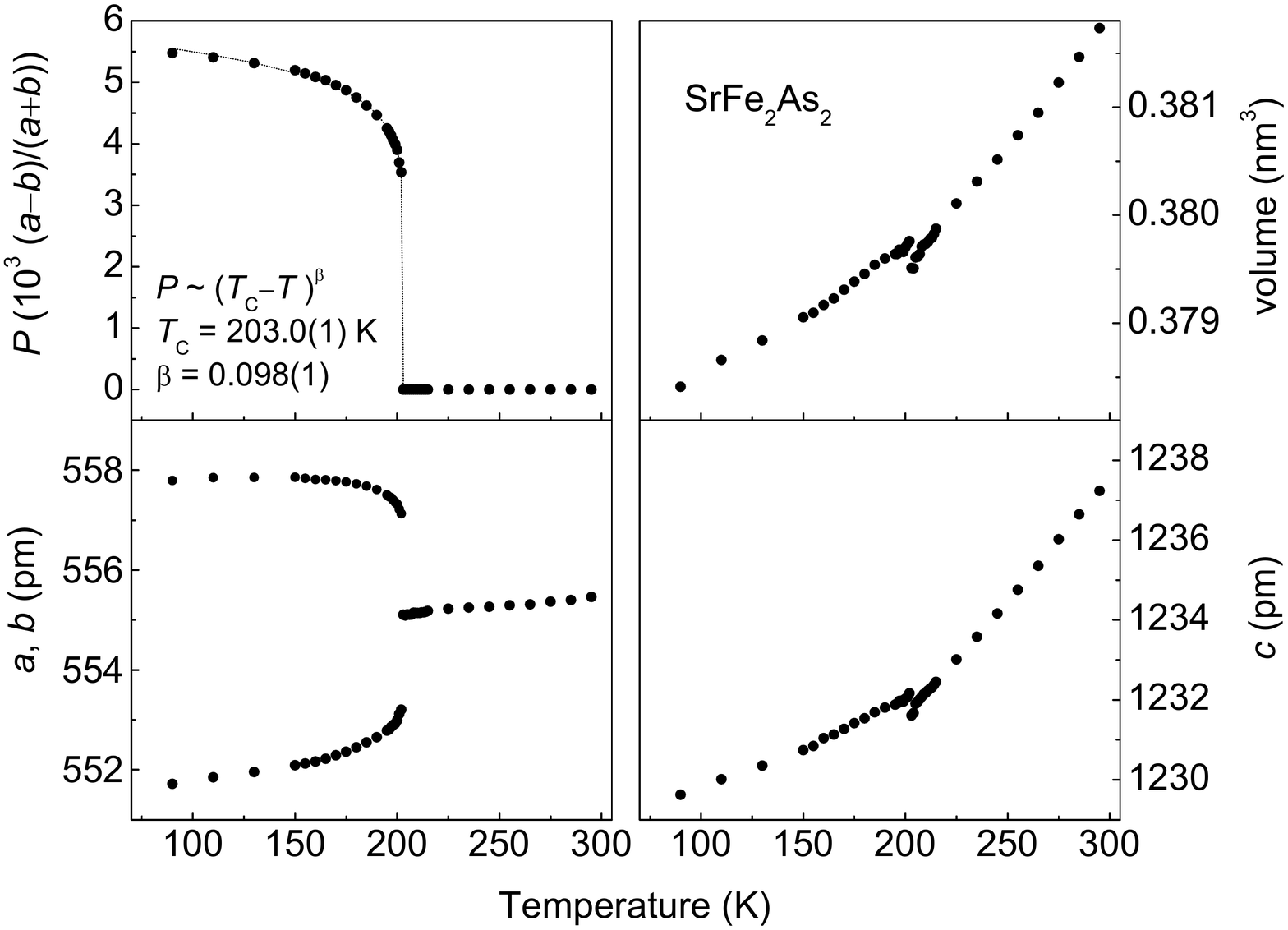}
\caption{\label{fig:SrFe2As2_LAT} Lattice parameters, cell volume and order parameter of \SFA. The line describes the power law fit. Error bars are within data points.}
}
\end{figure}

Figure~\ref{fig:SrFe2As2_LAT} shows the unit cell parameters of \SFA\ determined by Rietveld refinements. The $a$ lattice parameter in the tetragonal structure has been multiplied by $\sqrt2$ for comparison. We have found a rather abrupt splitting of the lattice parameters on cooling below 203 K. The tetragonal axis $a_{\mathrm{t}} = 555.11$~pm splits by +2.0~pm (+0.365\%) and $-1.9$~pm $(-0.343\%)$ within 1 K, leading to $a_{\mathrm{o}} = 557.13(3)$~pm and $b_{\mathrm{o}} = 553.20(3)$~pm at 202 K, respectively. Below this temperature, we observe a continuous increase of $a_{\mathrm{o}}$ and decrease of $b_{\mathrm{o}}$. The $a_{\mathrm{o}}$ parameter saturates already at 165 K towards a total change of +0.67\%, whereas ($b_{\mathrm{o}}$) decreases further down to a total change of $-0.94\%$ at 90 K. The different behavior of the orthorhombic axis is most likely a consequence of the magnetic ordering involved in the structural transition. We can understand this anisotropy, if we assume that the spin moments at the iron atoms in \SFA\ are aligned parallel to \textit{one} orthorhombic axis and perpendicular to the other one, as known for LaFeAsO \cite{Cruz-Neutrons} and recently reported for \BFA\ \cite{Huang-2008-BFA-Neutron}.

The temperature dependence of the order parameter $P = \frac{a-b}{a+b}$ is presented in Figure~\ref{fig:SrFe2As2_LAT}. We can fit the data to a power law $P = q(\frac{T_{tr}-T}{T})^\beta$ and obtain a critical exponent $\beta$ = 0.098(1) and a critical temperature $T_{tr}$ = 203.0(1) K. Simple power-law dependencies defining critical exponents are generally valid close to the critical point, but the values of the critical exponents are often very different from the Landau prediction ($\beta = \frac{1}{2})$. However, it is well-known that the Landau theory fails to provide a general description of critical phenomena. In the present case, the two-dimensional character of the structure has to be taken into account and the exponent $\beta \approx 0.1$ is not too far from $\frac{1}{8}$, which is the prediction of the 2-dimensional Ising model \cite{Lines-1977, Stanley-1971}.

So far, the structural transition of \SFA\ has a clear signature of second order, but on the other hand, some results point to a first order mechanism. Typical signs for first order transitions are the occurence of hysteresis and a volume jump at $T_{tr}$. Indeed, specific heat measurements \cite{Krellner-2008} show a hysteresis of $\approx$ 0.2 K. This value is unexpectedly small, but quite consistent with our results regarding the cell volume. We find a jump at $T_{tr}$ as discernable from the volume plot in Figure~\ref{fig:SrFe2As2_LAT}, but the value $\frac{\Delta V}{V}$ is only 0.07\%. This is extremely small, especially if we keep in mind that the changes of the $a$ and $b$ lattice parameters are around 1\%.  We do not believe that these very small magnitudes of hysteresis and volume jump clearly justify a first order transition mechanism.

\hspace{2cm}

\begin{figure}[h]
\center{
\includegraphics[width=0.7\textwidth]{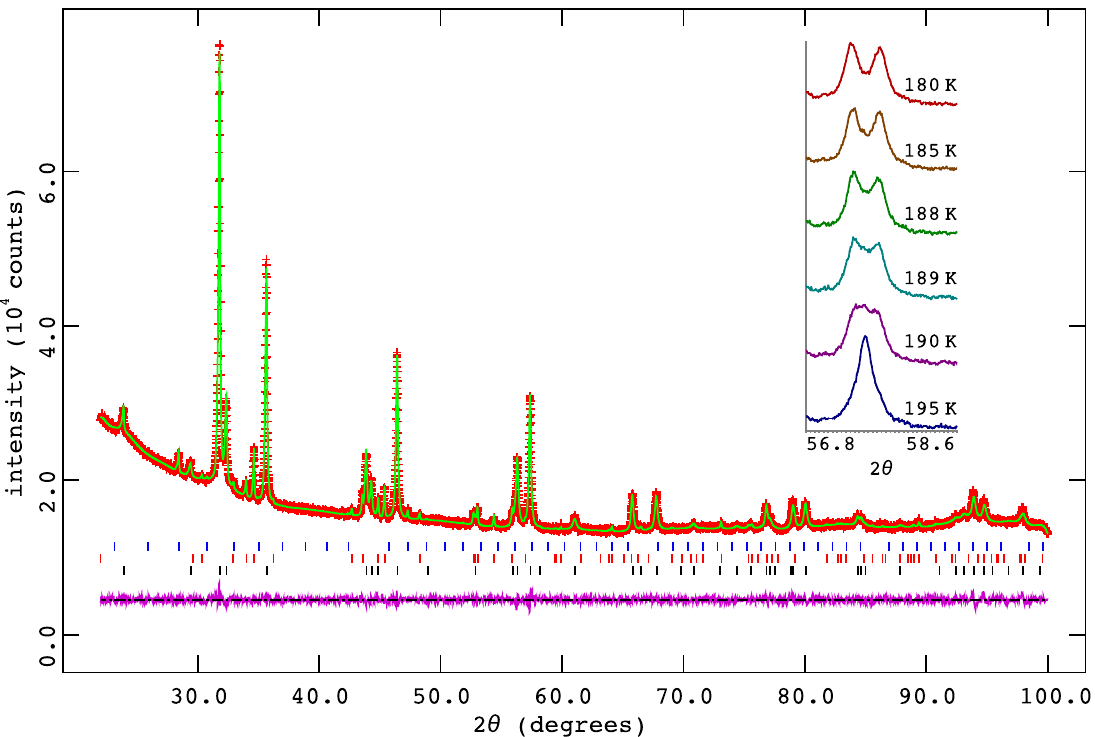}
\caption{\label{fig:EuFe2As2_X} (Color online) X-ray powder pattern mesured at 297 K (+) and Rietveld fit (-) of \EFA. FeAs (\textless 20\%) and Eu$_2$O$_3$ (\textless 1\%) were included as minor impurity phases. Insert: Splitting of the (2 1 3) reflection.}
}
\end{figure}

\hspace{2cm}

\begin{figure}[h]
\center{
\includegraphics[width=0.7\textwidth]{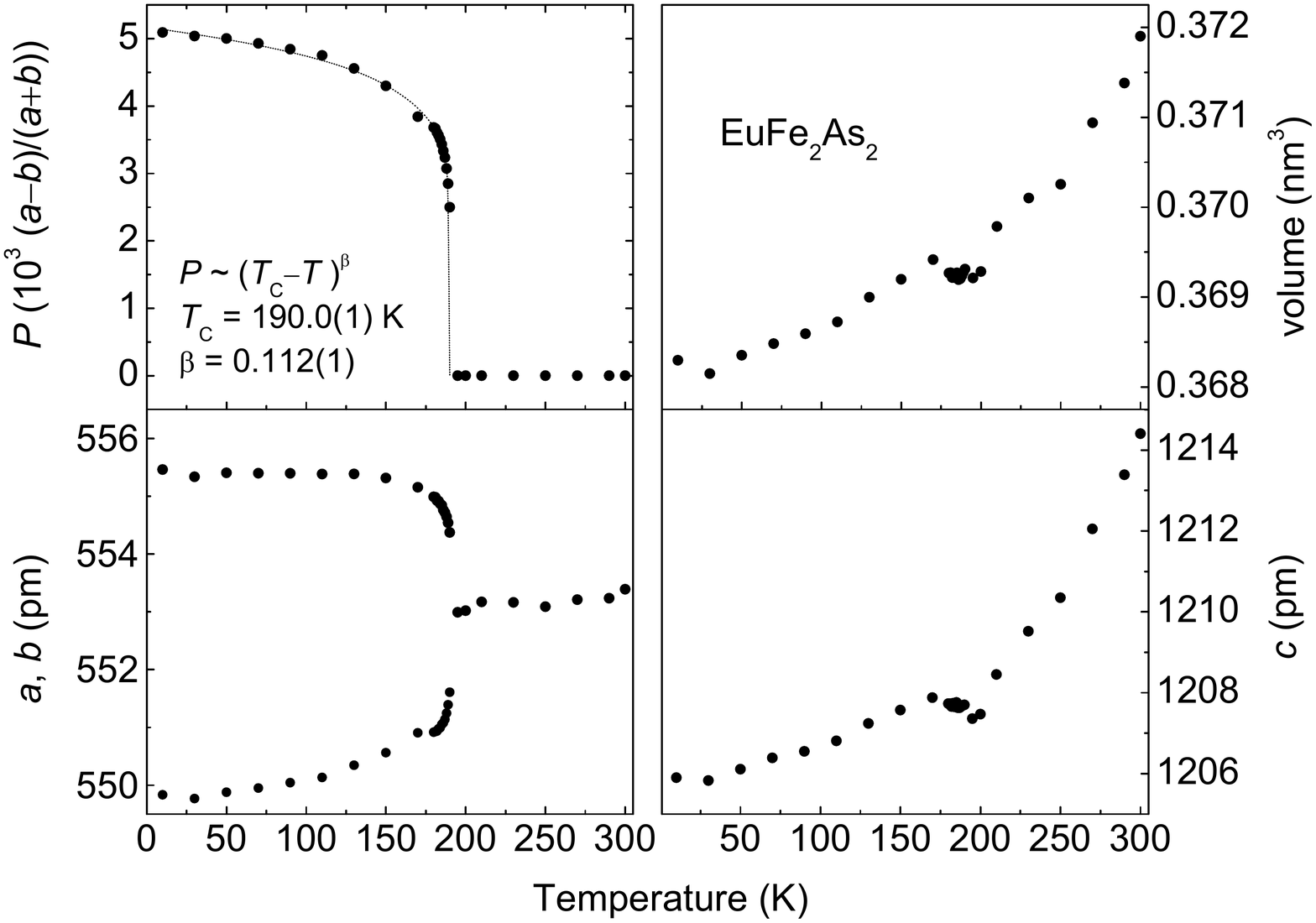}
\caption{\label{fig:EuFe2As2_LAT} Lattice parameters, cell volume and order parameter of \EFA. The line describes the power law fit. Error bars are within data points.}
}
\end{figure}

This point of view is supported by the comparison of the structural data of \SFA\ with other \AFA\ compounds. Figure~\ref{fig:EuFe2As2_X} shows the Rietveld fit of the \EFA\ sample. Crystallographic data of the ambient and low-temperature structures are compiled in Tables~\ref{tab:CrystallographicLT} and \ref{tab:CrystallographicRT}. In Figure~\ref{fig:EuFe2As2_LAT} we present the lattice and order parameter of \EFA. We were not able to determine the exact progression of the $c$ axis and thus the volume in the proximity of $T_{tr}$. A behavior very similar to \SFA\ is observed, but the $T_{tr}$ decreases to 190.0(1) K and the critical exponent increases to $\beta$ = 0.112(1). This trend continues in the case of \BFA\ with $T_{tr}$  = 139 K and $\beta$ = 0.142 \cite{BFA}. Obviously, the critical exponent $\beta$ scales with $T_{tr}$ and the transition indeed converges towards first order (where $\beta$ vanishes), but remains second order even in \SFA. This is also what to expect from group theory. It is worth mentioning, that the abruptness of a phase transition is by no means a cast-iron argument for a first order mechanism. This is, however, a question of data accuracy close to the critical point.

\section{Conclusion}

In summary, our results suggest a complex nature of the phase transitions in \AFA\ compounds due to several competing order parameters with respect to its structural and magnetic components. We have shown by M\"ossbauer spectroscopy that the transition of \SFA\ is accompanied by magnetic ordering as it is the case for the Ba and Eu compounds. We are aware that the magnetic ordering couples to the lattice and that both contributions are hard to distinguish due to their proximity. However, the focus in the current work has been on the structural part. Our precise determination of the lattice parameters close to $T_{tr}$ are indicative of a second order transition with continuous varying order parameters and simple power-law dependencies for both \SFA\ and \EFA. This is typical for displacive structural transitions and consistent with the group-subgroup relationship between the $I4/mmm$ and $Fmmm$ space groups. The comparison between the data of \SFA, \EFA, and \BFA\ clearly reveals a relation between the transition temperatures and the critical exponents. Obviously, the transition becomes more and more continuous as $T_{tr}$ decreases from \SFA\ towards \BFA. But of course, the mechanism is the same in all three cases. From this we conclude that all \AFA\ compounds with $A$ = Ba, Sr, and Eu undergo second order displacive structural transitions.



\section*{References}

\bibliographystyle{unsrt}


\begin{table}[p]
\caption{\label{tab:CrystallographicLT} Crystallographic data of low-temperature \SFA\ and \EFA.}

\begin{center}
\begin{tabular}{lll}
                 &      &      \\
                 & \SFA & \EFA \\
 Temperature (K) & 90   & 10 \\
 Space group & $Fmmm$           & $Fmmm$ \\
 \textit{a} (pm) & 557.83(3)        & 555.46(2) \\
 \textit{b} (pm) & 551.75(3)        & 549.83(2) \\
 \textit{c} (pm) & 1229.65(6)       & 1205.90(4) \\
 \textit{V} (nm$^{3}$) & 0.37846(1) & 0.36830(4) \\
 \textit{Z} & 4                   & 4 \\
 $\rho_{calc}$ (g/cm$^3$) & 6.11  & 7.46 \\
 data points & 5823               & 7650 \\
 reflections (all phases) & 145        & 306 \\
 atomic parameters & 5         & 5 \\
 profile parameters & 4         & 10 \\
 background parameters & 36       & 36 \\
 other parameters & 12            & 24 \\
 \textit {d} range & $0.963 - 6.148$ & $0.945 - 7.681$ \\
 R$_{P}$, \textit{w}R$_{P}$ & 0.0361, 0.0487 & 0.0131, 0.0237\\
 R$(F2)$, $\chi2$ & 0.104, 4.444 & 0.064, 2.371\\
 GooF & 2.11 & 1.54 \\
Atomic parameters: \\
 Sr,Eu & 4$a$ (0,0,0)                               &  4$a$ (0,0,0)\\
    & $U_{iso} = 180(40)$                         & $U_{iso} = 73(6)$            \\
 Fe & 8$f$ ($\frac{1}{4},\frac{1}{4},\frac{1}{4}$)   &  8$f$ ($\frac{1}{4},\frac{1}{4},\frac{1}{4}$)\\
    & $U_{iso} = 60(20)$                           & $U_{iso} = 47(6)$\\
 As & 8$i$ (0,0,$z$)                                 &  8$i$ (0,0,$z$)  \\
    & $z$ = 0.3612(3)                                &  $z$ = 0.3632(1) \\
    & $U_{iso} = 60(20)$                          & $U_{iso} = 87(6)$\\
Bond lengths (pm):\\
(Sr,Eu)--As  &  324.4(3)$\times$4  & 320.6(1)$\times$4\\
   & 327.0(3)$\times$4 & 323.0(1)$\times$4\\
Fe--As  &  239.1(3)$\times$4          & 238.3(1)$\times$4\\
Fe--Fe  &  278.9(1)$\times$2    & 277.7(1)$\times$2 \\
   & 275.9(1)$\times$2 & 274.9(1)$\times$2 \\
Bond angles (deg):\\
As--Fe--As &  110.2(2)$\times$2                             & 110.1(1)$\times$2\\
           &  109.5(1)$\times$2   & 109.6(1)$\times$2\\
    &108.6(1)$\times$2  & 108.7(1)$\times$2\\
\end{tabular}
\end{center}
\end{table}

\begin{table}[p]
\caption{\label{tab:CrystallographicRT} Crystallographic data of \SFA\ and \EFA\ at 297K.}

\begin{center}
\begin{tabular}{lll}
                 &      &      \\
                 & \SFA & \EFA \\
 Space group & $I4/mmm$           & $I4/mmm$ \\
 \textit{a} (pm) & 392.43(1)        & 390.62(1) \\
 \textit{b} (pm) & $=a$        & $=a$ \\
 \textit{c} (pm) & 1236.44(1)       & 1212.47(2) \\
 \textit{V} (nm$^{3}$) & 0.19041(1) & 0.18501(1) \\
 \textit{Z} & 2                   & 2 \\
 $\rho_{calc}$ (g/cm$^3$) & 6.09  & 7.42 \\
 data points & 9000               & 7800 \\
 reflections (all phases) & 46        & 281 \\
 atomic parameters & 4         & 4 \\
 profile parameters & 8         & 8 \\
 background parameters & 36       & 36 \\
 other parameters & 5            & 20 \\
 \textit {d} range & $0.981 - 6.182$ & $0.960 - 7.678$ \\
 R$_{P}$, \textit{w}R$_{P}$ & 0.0232, 0.0306 & 0.0147, 0.0209\\
 R$(F2)$, $\chi2$ & 0.029, 1.601 & 0.047, 2.580\\
 GooF & 1.27 & 1.61 \\
Atomic parameters: \\
Sr, Eu & 2$a$ (0,0,0)                         &  2$a$ (0,0,0)\\
    & $U_{iso} = 129(5)$                    & $U_{iso} = 123(5)$            \\
 Fe & 4$d$ ($\frac{1}{2},0,\frac{1}{4}$)   &  4$d$ ($\frac{1}{2},0,\frac{1}{4}$)\\
    & $U_{iso} = 72(4)$                   & $U_{iso} = 91(6)$\\
 As & 4$e$ (0,0,$z$)                       &  4$e$ (0,0,$z$)  \\
    & $z$ = 0.3600(1)                    &  $z$ = 0.3625(1) \\
    & $U_{iso} = 86(4)$                  & $U_{iso} = 96(5)$\\
Bond lengths (pm):\\
(Sr,Eu)--As  &  327.0(1)$\times$8     & 322.6(1)$\times$8\\
Fe--As  &  238.8(1)$\times$4          & 238.2(1)$\times$4\\
Fe--Fe  &  277.5(1)$\times$4    & 276.2(1)$\times$2 \\
Bond angles (deg):\\
As--Fe--As &  110.5(1)$\times$2                             & 110.1(1)$\times$2\\
     &108.9(1)$\times$4  & 109.1(1)$\times$2\\
\end{tabular}
\end{center}
\end{table}

\end{document}